\documentclass[english,5p,letterpaper]{elsarticle}
\usepackage[T1]{fontenc}
\usepackage[latin9]{inputenc}
\usepackage{babel}
\usepackage{array}
\usepackage{amssymb}
\usepackage{graphicx}
\usepackage[unicode=true,
 bookmarks=true,bookmarksnumbered=false,bookmarksopen=false,
 breaklinks=false,pdfborder={0 0 1},backref=false,colorlinks=false]
 {hyperref}
\usepackage{breakurl}

\makeatletter

\providecommand{\tabularnewline}{\\}

\usepackage{sidecap} % added by sidecap.module

\usepackage{natbib}
\usepackage{multicol}
\usepackage{lineno}

\@ifundefined{showcaptionsetup}{}{%
 \PassOptionsToPackage{caption=false}{subfig}}
\usepackage{subfig}
\makeatother

\begin{document}

\title{Data Acquisition and Readout System for the LUX Dark Matter Experiment}

\author[cwru]{D.\,S. Akerib}

\author[sdsmt]{X. Bai}

\author[yale]{S. Bedikian}

\author[yale]{E. Bernard}

\author[llnl]{A. Bernstein}

\author[cwru]{A. Bradley}

\author[yale]{S.\,B. Cahn}

\author[cwru]{M.\,C. Carmona-Benitez}

\author[llnl]{D. Carr}

\author[brown]{J.\,J. Chapman\corref{cor1}}

\cortext[cor1]{Corresponding Author: Jeremy\_Chapman@brown.edu}

\author[cwru]{K. Clark}

\author[davis]{T. Classen}

\author[cwru]{T. Coffey}

\author[yale]{A. Curioni}

\author[llnl]{S. Dazeley}

\author[brown]{L. de\,Viveiros}

\author[cwru]{M. Dragowsky}

\author[uor]{E. Druszkiewicz}

\author[brown]{C.\,H. Faham}

\author[brown]{S. Fiorucci}

\author[brown]{R.\,J. Gaitskell}

\author[cwru]{K.\,R. Gibson}

\author[umd]{C. Hall}

\author[sdsmt]{M. Hanhardt}

\author[davis]{B. Holbrook}

\author[ucb]{M. Ihm}

\author[ucb]{R.\,G. Jacobsen}

\author[yale]{L. Kastens}

\author[llnl]{K. Kazkaz}

\author[davis]{R. Lander}

\author[yale]{N. Larsen}

\author[cwru]{C. Lee}

\author[umd]{D. Leonard}

\author[lbl]{K. Lesko}

\author[yale]{A. Lyashenko}

\author[brown]{D.\,C. Malling}

\author[tamu]{R. Mannino}

\author[yale]{D.\,N. McKinsey}

\author[usd]{D. Mei}

\author[davis]{J. Mock}

\author[harvard]{M. Morii}

\author[ucsb]{H. Nelson}

\author[yale]{J.\,A. Nikkel}

\author[brown]{M. Pangilinan}

\author[cwru]{P. Phelps}

\author[cwru]{T. Shutt}

\author[uor]{W. Skulski}

\author[llnl]{P. Sorensen}

\author[usd]{J. Spaans}

\author[tamu]{T. Stiegler}

\author[davis]{R. Svoboda}

\author[davis]{M. Sweany}

\author[davis]{M. Szydagis}

\author[davis]{J. Thomson}

\author[davis]{M. Tripathi}

\author[brown]{J.\,R. Verbus}

\author[davis]{N. Walsh}

\author[tamu]{R. Webb}

\author[tamu]{J.\,T. White}

\author[harvard]{M. Wlasenko}

\author[uor]{F.\,L.\,H. Wolfs}

\author[davis]{M. Woods}

\author[usd]{C. Zhang}

\address[brown]{Brown University, Dept. of Physics, 182 Hope St., Providence, RI
02912}

\address[cwru]{Case Western Reserve University, Dept. of Physics, 10900 Euclid
Ave, Cleveland, OH 44106}

\address[harvard]{Harvard University, Dept. of Physics, 17 Oxford St., Cambridge,
MA 02138}

\address[lbl]{Lawrence Berkeley National Laboratory, 1 Cyclotron Rd., Berkeley,
CA 94720}

\address[llnl]{Lawrence Livermore National Laboratory, 7000 East Ave., Livermore,
CA 94551}

\address{}

\address[sdsmt]{South Dakota School of Mines and Technology, 501 East St Joseph
St., Rapid City, SD 57701}

\address[tamu]{Texas A \& M University, Dept. of Physics, College Station, TX 77843}

\address[ucb]{University of California Berkeley, Dept. of Physics, Berkeley, CA
94720-7300}

\address[davis]{University of California Davis, Dept. of Physics, One Shields Ave.,
Davis, CA 95616}

\address[ucsb]{University of California Santa Barbara, Dept. of Physics, Santa
Barbara, CA 93106}

\address[umd]{University of Maryland, Dept. of Physics, College Park, MD 20742}

\address[uor]{University of Rochester, Dept. of Physics and Astronomy, Rochester,
NY 14627}

\address[usd]{University of South Dakota, Dept. of Physics, 414E Clark St., Vermillion,
SD 57069}

\address[yale]{Yale University, Dept. of Physics, 217 Prospect St., New Haven,
CT 06511}
\begin{abstract}
LUX is a two-phase (liquid/gas) xenon time projection chamber designed
to detect nuclear recoils from interactions with dark matter particles.
Signals from the LUX detector are processed by custom-built analog
electronics which provide properly shaped signals for the trigger
and data acquisition (DAQ) systems. The DAQ is comprised of commercial
digitizers with firmware customized for the LUX experiment. Data acquisition
systems in rare-event searches must accommodate high rate and large
dynamic range during precision calibrations involving radioactive
sources, while also delivering low threshold for maximum sensitivity.
The LUX DAQ meets these challenges using real-time baseline suppression
that allows for a maximum event acquisition rate in excess of 1.5~kHz
with virtually no deadtime. This paper describes the LUX DAQ and the
novel acquisition techniques employed in the LUX experiment. 
\end{abstract}
\maketitle

%\linenumbers

\section{Introduction}

The LUX dark matter experiment is a two-phase xenon Time Projection
Chamber (TPC). The goal of LUX is to detect energy depositions caused
by galactic dark matter particles, known as Weakly Interacting Massive
Particles or WIMPs, scattering off atomic nuclei \citep{Akimov200746,Alner2007287,Angle:2007uj,Aprile2005289}. 

\begin{figure*}
\includegraphics[width=2\columnwidth]{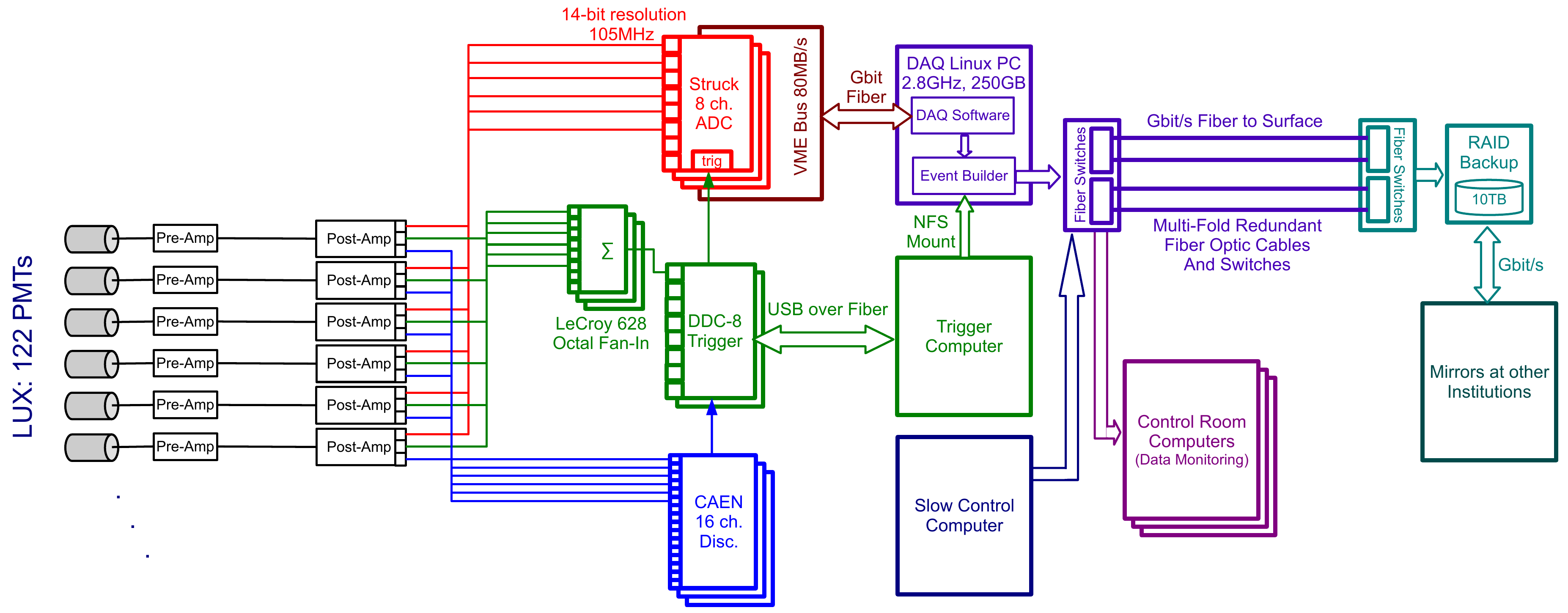}

\caption{\label{fig:DAQdiagram}Data flow diagram of the LUX electronics and
DAQ. Signals from the PMTs pass through air-side preamplifiers immediately
after exiting the xenon space. A postamplifier further amplifies and
shapes the signals. The postamplifier generates three outputs: one
for the Struck ADCs, one for the DDC-8 digital trigger system, and
one for the CAEN discriminators. The data from the Struck ADCs is
read out from the VME bus to the DAQ computer, where it is sorted
and prepared for analysis.}

\end{figure*}
The site for LUX is the Davis Laboratory at the $4850\,\mbox{ft.}$
level of the Sanford Underground Laboratory at the Homestake Mine
in Lead, SD. The detector is housed in an $8\,\mbox{m}$ diameter
($300\,\mbox{tonne}$) water tank that acts as a background shield
and muon veto. The LUX detector uses $350\,\mbox{kg}$ of liquid xenon,
$300\,\mbox{kg}$ of which define the active region. This active region
is observed by 122 Hamamatsu R8778 $2\,\mbox{inch}$ diameter photomultiplier
tubes (PMTs). There are 61 PMTs in the top array and 61 in the bottom
array. The PMTs have an average quantum efficiency (QE) of $33\pm2.3\%$
at the xenon scintillation wavelength of $175\,\mbox{nm}$ \citep{Hamamatsu}.
The active region is walled by PTFE (polytetrafluoroethylene) reflector
panels and is $59\,\mbox{cm in height}\times49\,\mbox{cm}$ in diameter
\citep{McKinseyTAUP2009,FiorucciSUSY2009}. Particle interactions
in the xenon create scintillation light and ionization electrons.
Wire grids at the top and bottom of the liquid provide a uniform electric
field in the active region. This electric field causes the electrons
to drift from the interaction site towards the liquid surface. Additional
grids at the surface provide a stronger electric field that extracts
the electrons from the liquid and accelerates them in xenon gas, creating
electro-luminescence. The measurement of the drift time and the PMT
hit pattern from the electro-luminescence allow for 3D localization
of the original particle interaction.

The LUX data acquisition system (DAQ) is designed to simultaneously
read out all 122 LUX PMTs, as well as the 20 PMTs observing the water
shield, summed into eight DAQ channels. The LUX readout is comprised
of a custom-built analog electronics chain and triggering system,
Struck fast 14-bit ADCs, and CAEN discriminators and scalers, as shown
in Fig. \ref{fig:DAQdiagram} \citep{struck,CAEN}. The Struck digitizer
firmware has been developed by the manufacturer in collaboration with
LUX to perform real-time baseline suppression. This paper presents
a description of the signals generated by interactions in the detector,
the analog electronics chain designed to read out the PMT signals,
and the data acquisition system. Data formatting and optimization
techniques are also described.

\section{Signals}

There are two types of interactions in the detector: nuclear recoils
(denoted by the \emph{r} subscript), which are interactions with the
xenon nucleus, and electron recoils (denoted by the \emph{ee} subscript),
in which the particle interacts with the electrons. A dark matter
WIMP particle is expected to produce a nuclear recoil in the detector,
whereas gamma ray interactions will produce electron recoils \citep{Gaitskell:2004rev}.
A particle interaction in the detector is characterized by two scintillation
signals. The first signal is the prompt scintillation of the interaction
in the liquid xenon, and is known as the S1 pulse. For interactions
in the xenon active region, S1 light is evenly distributed on the
PMT arrays in $xy$. The majority of the light (80\%) falls on the
bottom PMT array due to total internal reflection at the liquid surface.
For an energy deposition of $2\,\mbox{ke\ensuremath{\mbox{V}_{ee}}}$
in the center of the detector in the presence of a $1\,\mbox{kV/cm}$
electric field, the S1 pulse will be comprised of about 50~photons
\citep{Doke2002JJAP}. The fundamental rise-time of the S1 light signal
is limited by the response of the PMTs and is $\sim6\,\mbox{ns}$.
The S1 signal decays exponentially with an effective time constant
of $\tau=29\,\mbox{ns}$ \citep{Sorensen2008thesis,deViveiros2009thesis}.
The S1 pulses are fully characterized by the DAQ in a time window
of $<200\,\mbox{ns}$. Monte Carlo simulations using Geant4 indicate
that the S1 light collection in LUX will be $10\,\mbox{photoelectrons/ke\ensuremath{\mbox{V}_{ee}}}$
(at an energy of $122\,\mbox{ke\ensuremath{\mbox{V}_{ee}}}$) with
no electric field applied, and assuming 33\% PMT QE and 95\% PTFE
reflectivity in liquid xenon. The S1 light collection becomes $5\,\mbox{photoelectrons/ke\ensuremath{\mbox{V}_{ee}}}$
in the active region in the presence of an electric field of $1\,\mbox{kV/cm}$.

\begin{figure}[h]
\includegraphics[width=1\columnwidth]{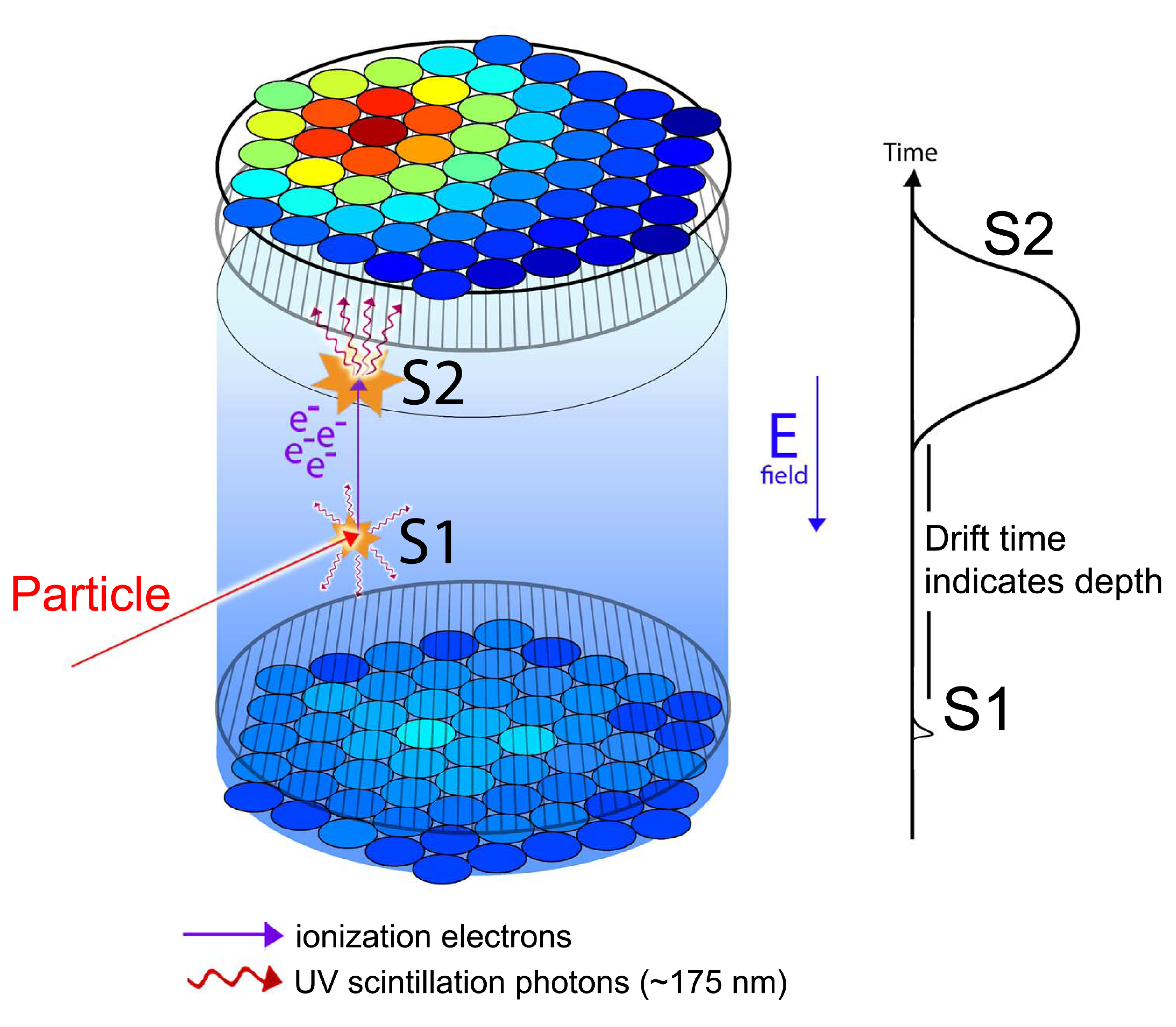}

\caption{\label{fig:LUXdiagram}Diagram of a particle interaction in the LUX
detector showing primary (S1) and secondary (S2) scinitllation signals.
The PMT hit pattern shown is generated by the S2 signal, where the
gray scale from white (least) to black (most) indicates the number
of photoelectrons detected by that PMT. }

\end{figure}

The initial particle interaction also ionizes xenon atoms, releasing
electrons at the interaction site. These electrons can recombine with
their host atoms, producing recombination light which increases the
size of the S1 signal. However, in LUX, the large electric field ($\gtrsim0.1\,\mbox{kV/cm}$)
causes these electrons to drift from the interaction site towards
the liquid surface before they can efficiently recombine \citep{Dahl2009thesis}.
Grids, $0.5\,\mbox{cm}$ above and below the liquid surface, generate
strong electric fields of $5\,\mbox{kV/cm}$ in the liquid and $10\,\mbox{kV/cm}$
in the gas \citep{McKinseyTAUP2009}. This causes the electrons to
leave the liquid and accelerate in the gas phase \citep{Bolozdynya1999314}.
The electrons produces further scintillation through electro-luminescence,
proportional to the number of electrons extracted (see Fig. \ref{fig:LUXdiagram}).
This is known as the S2 pulse, and is about three orders of magnitude
larger than the S1 pulse, depending on grid separation, applied field,
and xenon gas pressure \citep{Bolozdynya1999314}. The ratio of S2
area to S1 area is used to discriminate between nuclear and electron
recoils \citep{Angle:2007uj,Sorensen2008thesis}. LUX expects an S2
light yield of 35~photoelectrons per electron ($e^{-}$) extracted
from the liquid and $45\,\mbox{\ensuremath{e^{-}}/ke\ensuremath{\mbox{V}_{ee}}}$
for a gamma interaction in the detector at an energy of $10\,\mbox{ke\ensuremath{\mbox{V}_{ee}}}$
\citep{Sorensen2008thesis,deViveiros2009thesis,Dahl2009thesis,PhysRevA.12.1771}.
This light yield accounts for geometric light collection in the gas
phase where the S2 is generated, detection efficiency of the PMTs,
and the number of scintillation photons produced per extracted electron.
The S2 pulse waveform is gaussian-shaped when summed across all PMT
channels. The S2 pulse width varies slightly as a function of depth
because the cloud of electrons diffuses as it drifts towards the liquid
surface \citep{Sorensen2008thesis,deViveiros2009thesis,2011sorensenelectrons}.
The S2 is fully characterized by the DAQ in a time window of $\sim3\,\mbox{\ensuremath{\mu}s}$.
The S2 is spread over most PMTs, but shows significant localization
in the top array above the electron extraction site. Monte Carlo simulations
indicate that up to 22\% of the S2 light is collected by a single
PMT in the top array. This allows for $xy$ localization of the event
with $\sigma=0.3\,\mbox{cm}$ precision. In LUX, the time between
S1 and S2 pulses, the electron drift time, is between $0$ and $230\,\mbox{\ensuremath{\mu}s}$,
given a field of $1\,\mbox{kV/cm}$ and a maximum drift distance of
$50\,\mbox{cm}$ \citep{McKinseyTAUP2009,Miller1968PhysRev}. This
drift time is used to determine the $z$ coordinate of the interaction. 

\begin{figure*}
\begin{centering}
\subfloat{\includegraphics{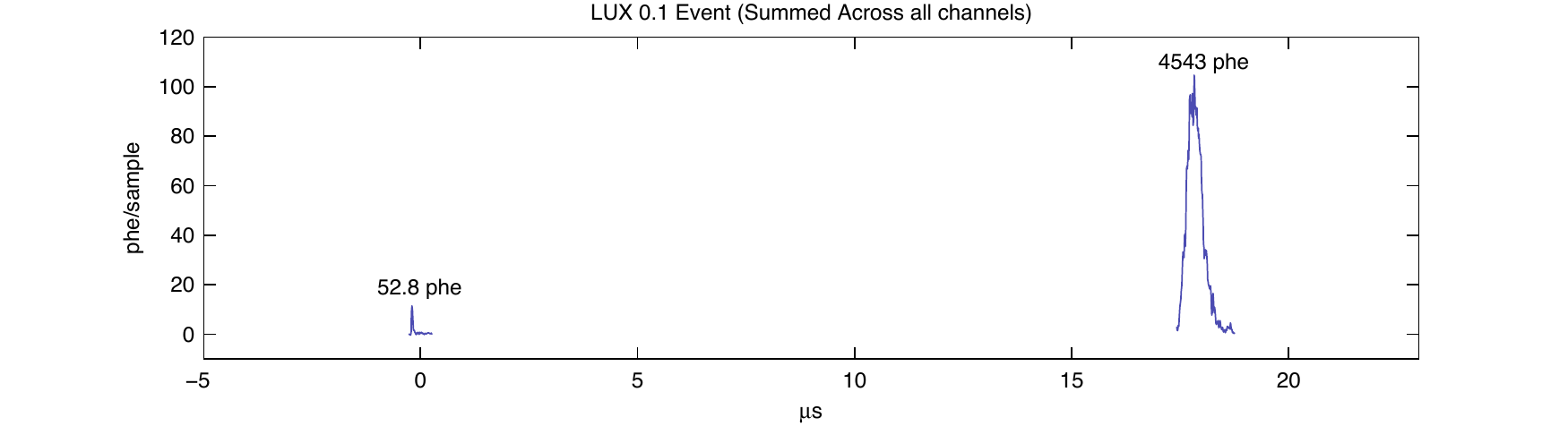}

}
\par\end{centering}

\begin{centering}
\subfloat{\includegraphics{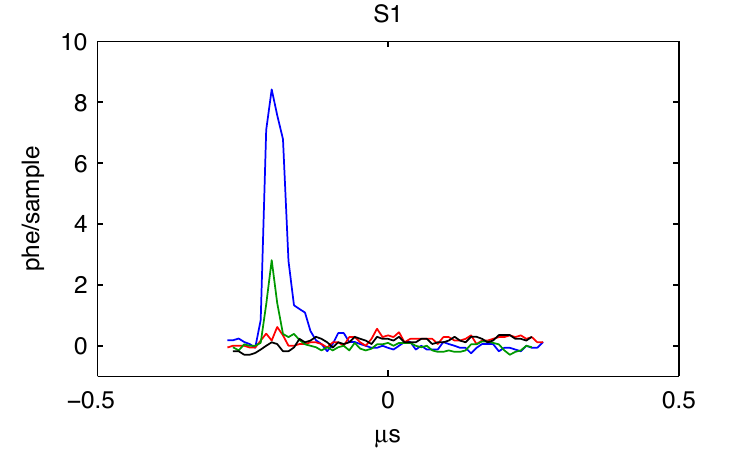}

}\hspace{0.6cm}\subfloat{\includegraphics{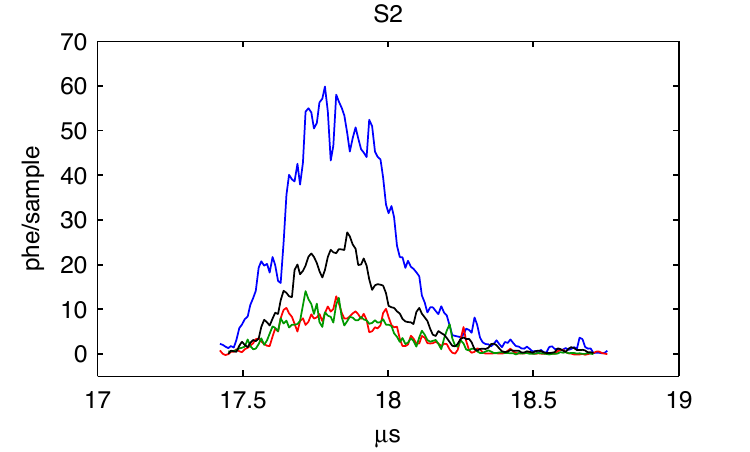}

}
\par\end{centering}

\caption{\label{fig:LUX-0.1-event}A $9.8\,\mbox{ke\ensuremath{\mbox{V}_{ee}}}$
event from LUX 0.1 using Pulse Only Digitization (Sect. \ref{sub:Pulse-Only-Digitization}).
(top) Sum of all four PMT channels. (bottom) The S1 and S2 signals
in the individual PMTs. The number of samples before and after each
pulse is selectable up to 24 and 31 respectively ($9.5\,\mbox{ns/sample}$).
Each PMT has a gain of $1\cdot10^{5}$. }

\end{figure*}

A typical event from LUX 0.1 is shown in Fig. \ref{fig:LUX-0.1-event}.
LUX 0.1 is a prototype detector using the LUX DAQ and electronics
with four PMTs (3 top, 1 bottom) \citep{FiorucciSUSY2009,PhelpsLUX0.1}.
The maximum drift time in LUX 0.1 is $25\,\mbox{\ensuremath{\mu}s}$.
The detector is quiet between the S1 and S2 pulses, which are the
most interesting features in the event.

\section{Electronics Chain}

The LUX PMT readout is comprised of both analog and digital electronics.
The analog electronics consists of two stages of amplifiers, designed
to shape the PMT signals for use by the triggering and acquisition
hardware.

\subsection{Digital Electronics\label{sec:Trigger-System}}

The LUX DAQ uses 16 Struck 8-channel fast ADC modules, model number
SIS3301, customized for LUX operation by Struck and Brown University.
The Struck ADCs digitize at $105\,\mbox{MHz}$ with a resolution of
14 bits per channel and an intrinsic noise of $20\,\mbox{nV/\ensuremath{\sqrt{Hz}}}$.
Each input has a total voltage range of $-1900\,\mbox{mV}\,\mbox{to}\,+100\,\mbox{mV}$.
Note that the PMT signals are negative, but for the purposes of this
paper the signals are described as positive. The Struck modules have
$2\times128\,\mbox{k}$ sample dual memory banks that allow for acquisition
on one bank while data is being downloaded from the other. The time
to switch memory banks is $17\,\mbox{\ensuremath{\mu}s}$. Each board
is connected to the VME bus, which is connected to the DAQ computer
via a fiber optic Gbit connection, using Struck communication modules
SIS3104 and SIS1100. The boards are capable of transferring data to
the acquisition computer at a maximum VME download speed of $80\,\mbox{MB/s}$
(2eVME protocol). Each board is controlled by four FPGAs, one per
pair of channels, whose firmware was developed by Struck in collaboration
with the Brown University group to operate in Pulse Only Digitization
(POD) mode, described in section \ref{sub:Pulse-Only-Digitization}.
Each Struck input has a $30\,\mbox{MHz}$ on-board single-pole anti-aliasing
filter. 

LUX is triggered using a digital triggering system, based on DDC-8DSP
designed and built for LUX by collaborators at the University of Rochester
and at SkuTek \citep{skutek}. The DDC-8 digital triggering system
digitizes the PMT signals at $64\,\mbox{MHz}$ with 14-bit resolution
for processing by the on-board FPGA. Each DDC-8DSP input has a $24\,\mbox{MHz}$
on-board anti-aliasing filter. The DDC-8 firmware has the ability
to identify S1 and S2 pulses. This allows the DDC-8 to trigger on
S1s, S2s, or events with both an S1 and an S2. The DDC-8 trigger
system has the ability to pre-select events of a given energy by measuring
the size of the S1. The hit pattern of the S2 in the top PMT array,
combined with the drift time between S1 and S2, allows the trigger
system to pre-select events of a given position. The LUX trigger system
and DDC-8DSP are discussed in \citep{LUXTrigger,skulski}.

\subsection{Analog Electronics}

The PMT readout electronics are designed to maximize signal to noise
in the dark matter search region of interest ($4-40\,\mbox{ke\ensuremath{\mbox{V}_{r}}}$)
and to allow for high-energy calibrations. The first two stages (see
Fig. \ref{fig:DAQdiagram}) in the analog electronics chain, the preamplifier
and the postamplifier, were designed and built by LUX collaborators
at U.\,C. Davis, Lawrence Livermore National Laborabory, and Harvard
University. The LUX readout requirements are based on the need for
low-energy calibrations in the dark matter search region, as well
as high-energy calibrations ($\gtrsim100\,\mbox{ke\ensuremath{\mbox{V}_{ee}}}$).
These calibrations are necessary to examine the detector response
to WIMP-like events, as well as to monitor the state of the detector
systems. The former requires that the PMT and amplifier gains be equal
to those used for WIMP searches. The latter allows for freedom in
varying the PMT and amplifier gains. Divided into these two categories,
there are four main detector calibrations: 

(i) Electron lifetime ($e$-lifetime) calibrations are used to measure
the purity of the xenon. Electro-negative impurities in the xenon
absorb drifting electrons. This causes the size of the S2 to vary
as a function of depth for a given interaction energy. The goal of
xenon purification is to minimize this effect \citep{Sorensen2008thesis,deViveiros2009thesis,PhelpsLUX0.1}.
The $e$-lifetime calibrations allow for a direct measurement of xenon
purity and the depth dependence of the S2 signal. This can be achieved
using gamma ray sources, such as $\mbox{\ensuremath{^{137}}Cs}$ or
$\mbox{\ensuremath{^{208}}Tl}$, which produce gamma rays of high
enough energies to penetrate into the center of the detector. 

(ii) Nuclear and (iii) electron recoil calibrations are used to measure
both relative scintillation efficiency and dark matter discrimination
efficiency between the two event types. Electron recoil events are
supplied using $\mbox{\ensuremath{^{137}}Cs}$, $\mbox{\ensuremath{^{208}}Tl}$,
$\mbox{\ensuremath{^{133}}Ba}$, and $\mbox{\ensuremath{^{57}}Co}$.
Nuclear recoil events are supplied using $\mbox{\ensuremath{^{252}}Cf}$
and $\mbox{AmBe}$ neutron sources. These calibrations must be made
with PMT and amplifier gains equal to those used for dark matter searches.

(iv) Internal calibrations are used to unambiguously measure the uniformity
of the detector. The two sources used for this are activated xenon
and $\mbox{\ensuremath{^{83m}}Kr}$ \citep{2009PhRvC..80d5809K}.

During dark matter search mode it is desirable to maximize trigger
efficiency at the lowest energy threshold, and to completely cover
the recoil energies expected for dark matter interactions. These requirements
demand careful balancing between low threshold and large dynamic range.

The analog electronics chain starts inside the detector at the PMT
base, which steps down the bias voltage for each dynode to produce
electron multiplication. Custom Gore coaxial cables transmit the signal
from the PMT base to the xenon space feedthrough \citep{gore}. A
single photoelectron pulse in this 10 meter long cable is attenuated
to 75\% of its original height while the pulse area is preserved.
On the air side, a preamplifier is mounted to the signal feedthrough.
The postamplifier is located near the Struck digitizers and the triggering
hardware. Between the preamplifier and postamplifier another 10 meter
long coaxial cable further attenuates the single photoelectron height
to 83\% of its original height, while the pulse area is preserved.
In order to meet the aforementioned specifications, $\gtrsim95\%$
of single photoelectron pulses should be clearly resolved from $5\sigma$
fluctuations in the baseline noise, and event energies up to $100\,\mbox{ke\ensuremath{\mbox{V}_{ee}}}$
should not saturate any stage of the electronics. The R8778 PMTs are
operated with a typical bias of $1.25\pm0.1\,\mbox{kV}$ to produce
a of gain $3.3\cdot10^{6}$. Higher energy calibrations that will
saturate the PMTs or electronics at this bias can be done by lowering
either the PMT or the amplifier gains. At a gain of $3.3\cdot10^{6}$
a single photoelectron generates a pulse area of $13.3\,\mbox{mV\ensuremath{\cdot}ns}$
at the output of the PMT base with $50\,\Omega$ termination. The
single photoelectron peak resolution is $\sigma=37\%$ on average
\citep{LUXPMTs}. The preamplifier has a gain of $\times5$, with
an in-line switchable attenuator of $\times1/10$, for an optional
total gain of $\times0.5$. The postamplifier has three outputs: one
for the Struck digitizers, one for the DDC-8 trigger system, and a
third, for the discriminator and scaler modules. The postamplifier
is an integrating amplifier, designed for an output bandwidth of $30\,\mbox{MHz}$,
in order to match the input bandwidth of the Struck and the DDC-8
digitizers. The postamplifier is designed to give a gain in pulse
area of $\times1.5$ for the Struck digitizers and $\times2.8$ for
the DDC-8 trigger system. The output for the discriminators has a
gain of $\times18$. Each of these outputs has optional $\times0.5$
attenuators. The Struck and the DDC-8 digitizers each have bandwidth
limiting filters at their inputs with cutoff frequencies of $30\,\mbox{MHz}$
and $24\,\mbox{MHz}$, respectively. Each PMT channel is digitized
by one Struck channel. The analog sum of eight PMT channels from the
DDC-8 output of the postamplifier is generated using LeCroy 628 octal
fan-in modules and sent to one channel of the DDC-8 trigger system.
This is done in order to reduce the cost of the trigger system. The
trigger system is discribed in more detail in Refs. \citep{LUXTrigger,skulski}.

\subsubsection{Dynamic Range:}

A design goal of the DAQ and electronics is that the maximum measureable
energy deposition in the detector be limited by the PMTs and not by
the electronics chain. The R8778 PMTs become 2\% nonlinear with an
instantaneous current draw of $13\,\mbox{mA}$ at the anode \citep{Hamamatsu,LUXPMTs}.
Simulations predict that a PMT in the top array, located directly
above the electron extraction site, will see at most 22\% of the total
S2 light. The maximum area of the S2 signal, combined across all channels
without saturating any PMT, is $2\cdot10^{5}\,\mbox{photoelectrons}$.
This corresponds to a maximum event energy of $\sim120\,\mbox{ke\ensuremath{\mbox{V}_{ee}}}$,
well above the desired calibration range of $100\,\mbox{ke\ensuremath{\mbox{V}_{ee}}}$.

At the Struck output of the postamplifier, a single photoelectron
has a height of $4\,\mbox{mV}$ and an area of $100\,\mbox{mV\ensuremath{\cdot}ns}$,
assuming a PMT gain of $3.3\cdot10^{6}$. The S1 light is typically
split; 20\% on the top PMT array and 80\% on the bottom. The S1 signal
on the bottom PMT array is evenly spread across all 61 channels. With
a maximum input voltage on the Struck ADC of $1900\,\mbox{mV}$, the
largest non-saturating S1 pulse has an area of $4.2\cdot10^{4}\,\mbox{photoelectrons}$.
This corresponds to a maximum energy deposition of $8.4\,\mbox{Me\ensuremath{\mbox{V}_{ee}}}$,
assuming a light collection efficiency of $5\,\mbox{photoelectrons/ke\ensuremath{\mbox{V}_{ee}}}$
in the presence of an electric field $\gtrsim0.5\,\mbox{kV/cm}$.
The maximum area of the S2 signal combined across all channels, without
saturating the Struck ADC in any single channel, is $1.1\cdot10^{6}\,\mbox{photoelectrons}$.
This corresponds to a maximum event energy of $715\,\mbox{ke\ensuremath{\mbox{V}_{ee}}}$.

\subsubsection{Threshold:}

The Struck SIS3301 is a 14-bit digitizer with a total voltage range
of $2\,\mbox{V}$.  A threshold of 13 ADC counts is below the height
of 95\% of single photoelectron signals. The total noise from the
electronics chain and ADC, as measured at the Struck input, is $155\,\mbox{\ensuremath{\mu}\ensuremath{\mbox{V}_{RMS}}}$.
A $5\sigma$ upward fluctuation of this noise is 6.5 ADC counts. This
is comfortably below the threshold of 13 ADC counts.

\section{Pulse Only Digitization\label{sub:Pulse-Only-Digitization}}

\begin{figure*}
\begin{centering}
\subfloat{\centering{}\includegraphics[width=2\columnwidth]{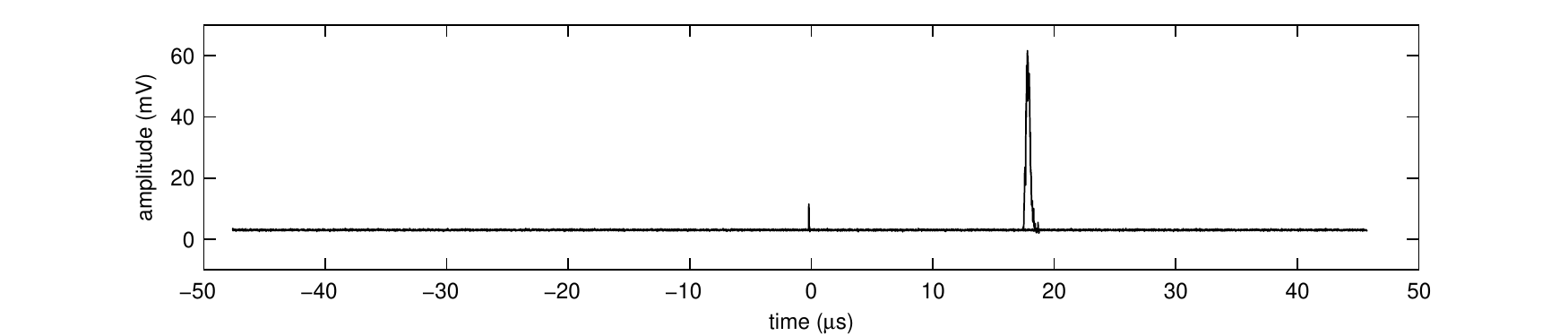}}
\par\end{centering}

\begin{centering}
\subfloat{\centering{}\includegraphics[width=2\columnwidth]{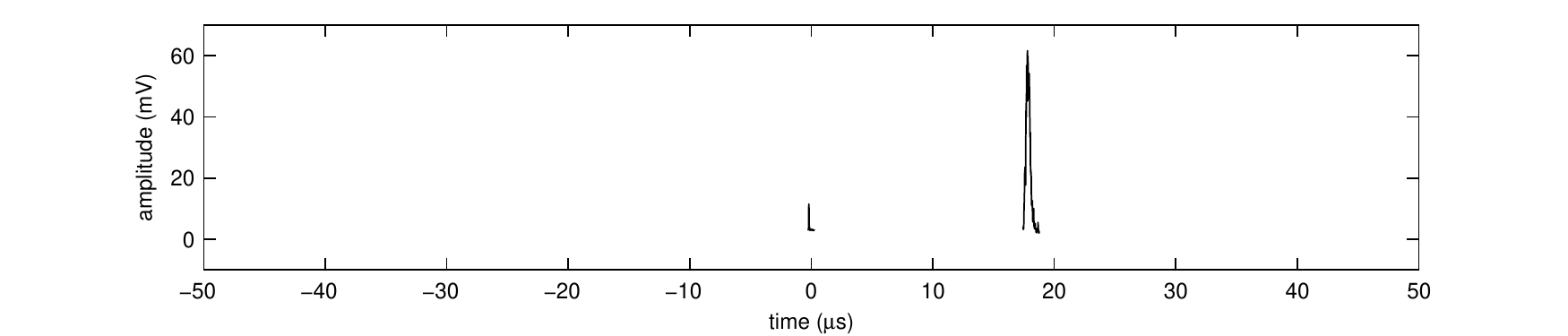}}
\par\end{centering}

\caption{\label{fig:PODexample}Sample event from the bottom PMT in LUX 0.1
digitized with (bottom) and without (top) Pulse Only Digitization
(POD).}

\end{figure*}

Pulse Only Digitization (POD) is an acquisition mode optimized for
signals that are dominated by long periods of baseline, interspersed
with short pulses. The purpose of POD mode is to acquire only the
pulses in the data stream, while suppressing the baseline. POD operates
by saving to memory only signals that rise above a threshold. This
improves the DAQ livetime by preventing the baseline between pulses
from dominating the data collected . An example of an event with and
without POD mode from LUX 0.1 is shown in Fig. \ref{fig:PODexample}. 

POD mode operates on channel pairs since the Struck memory banks are
organized per pair of channels. When one channel of a pair goes above
threshold and saves data to memory, the data from the partner channel
is recorded as well, even if the signal on that channel did not go
above threshold. 

Each pulse is recorded in the Struck memory with a header. This header
has a unique identifier to signify the beginning of a pulse in memory.
It also contains the 48-bit timestamp which is relative to the start
of the acquisition, the number of samples in the pulse, the value
of the baseline in ADC counts when the pulse was recorded, and a flag
indicating which channel(s) in the pair went above threshold. A separate
set of registers contain the timestamps and memory addresses of the
first 256 pulses saved to memory. This facilitates download of only
a portion of the data from the memory bank. This is used in single-event
mode to download only the pulses that are associated with the event
trigger, as discussed in Sect. \ref{sec:Acquisition-Strategies}.

\begin{figure}
\includegraphics[width=1\columnwidth]{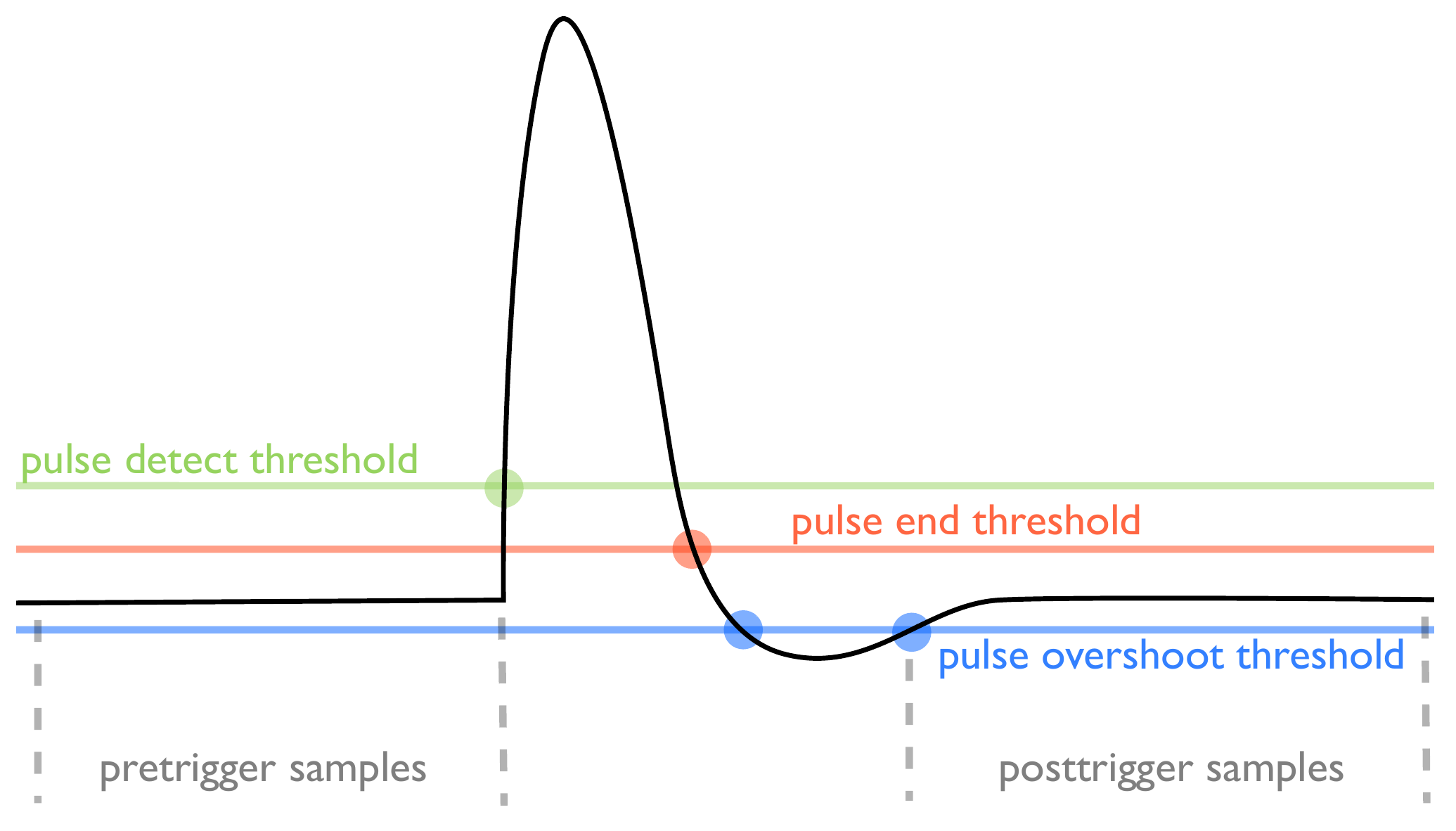}

\caption{\label{fig:PODdiagram}Illustration of Pulse Only Digitization (POD)
mode on an arbitrary signal (see Sect. \ref{sub:Pulse-Only-Digitization}).}

\end{figure}

Pulse acquisition on each channel is controlled by three independent
thresholds, set prior to the start of the acquisition. Each threshold
is illustrated in Fig. \ref{fig:PODdiagram}. The \emph{pulse detect
threshold} indicates the start of the pulse; \emph{pulse end threshold}
indicates the end of the pulse; and the \emph{pulse overshoot threshold}
can be used to record portions of the pulse that overshoot the baseline.
Pulse overshoot occurs when the signal crosses zero for a short time
before returning to the baseline. This feature ensures that bipolar
noise glitches are fully characterized by the DAQ for analysis. These
thresholds can be independently specified for each channel. 

\begin{figure}
\includegraphics[width=1\columnwidth]{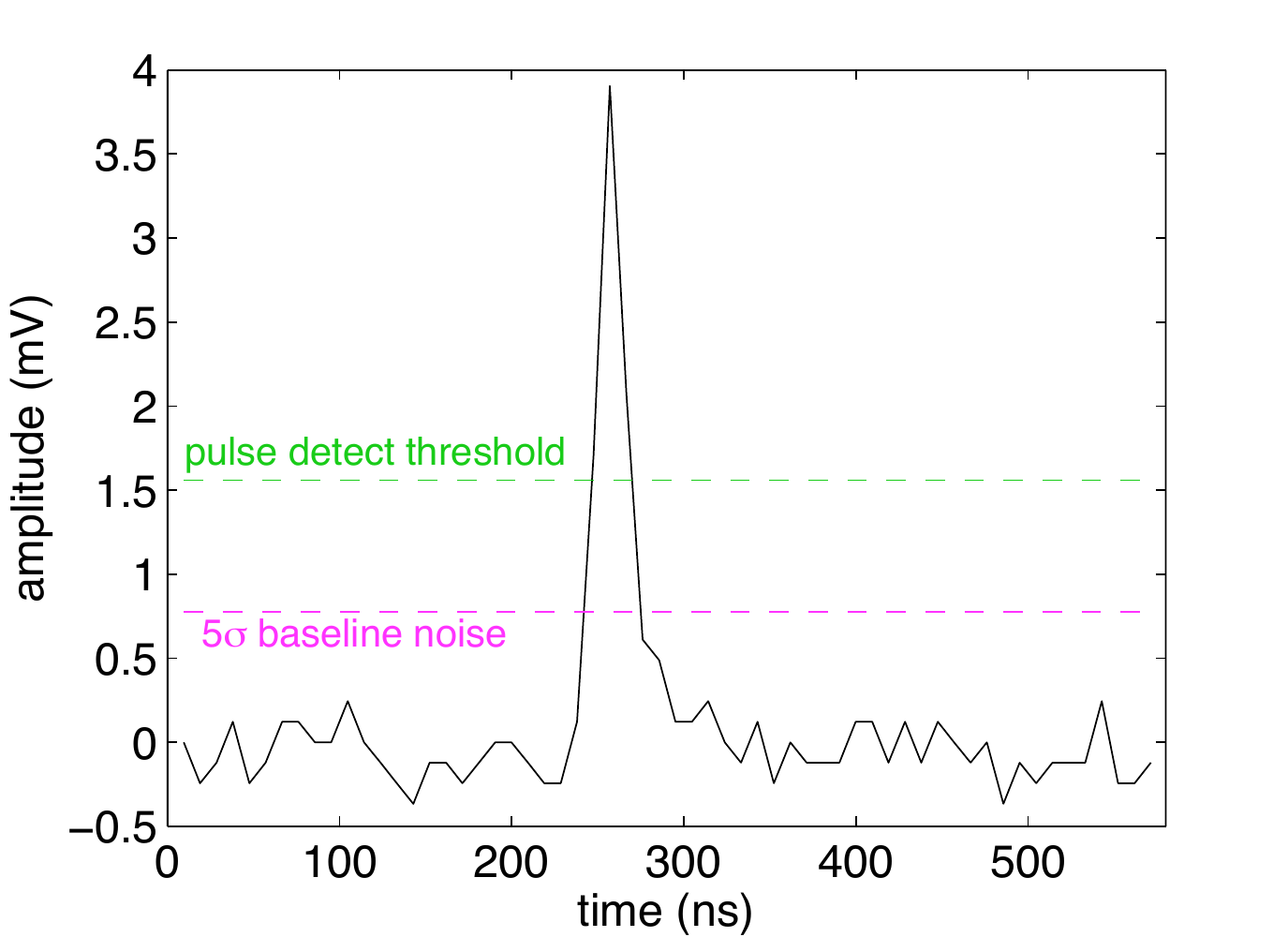}

\caption{\label{fig:sphe}Single photoelectron pulse from a PMT in LUX detected
with POD mode (see text for details). The \emph{pulse detect threshold}
corresponding to 95\% of single photoelectrons, and $5\sigma$ upward
fluctuation in the baseline noise are also indicated. }

\end{figure}

Each digitized sample is placed in a rolling buffer of 24 samples.
When a pulse is detected, up to 24 samples of pretrigger can be recorded
with it. Each pulse can have as many as 31 posttrigger samples recorded
as well. An example of a single photoelectron pulse, digitized in
LUX with POD mode, is shown in Fig. \ref{fig:sphe}. The \emph{pulse
detect threshold} is set at 13 ADC counts and corresponds to a single
photoelectron detection efficiency of 95\%. The $5\sigma$ upward
fluctuation in the baseline noise is also indicated. At this threshold,
with the maximum pretrigger and posttrigger samples, the baseline
suppression of the DAQ in POD mode is 99.99\%. This is measured by
keeping the PMT bias at $-100\,\mbox{V}$, below the threshold necessary
to genrate detectable signals, and counting the number of samples
recorded per second of acquisition. The only samples recorded are
those associated with noise glitches. This is addressed with the Valid
Pulse Trigger Gate, described in Sect. \ref{sub:Valid-Pulse-Trigger}.
POD mode gives a factor of $\times1/50$ reduction in storage size
for a typical event over the full drift length of LUX, compared to
non-POD mode acquisition.

The acquisition is designed to allow for slow changes in the baseline
caused by low-frequency noise, such as $60\,\mbox{Hz}$ pickup. The
baseline of each channel is determined by a rolling average of 16,
32, 64, or 128 samples. This number specified by the user at the start
of the acquisition. 128 samples was chosen for LUX, based on the noise
profile and the PMT signals. The pulse detection thresholds are measured
relative to this baseline. This rolling average is calculated until
the \emph{pulse detect threshold} is crossed, at which point its value
is recorded in the pulse header. By freezing the calculated baseline
at the beginning of the pulse, a large signal will not artificially
raise the rolling average, which would in turn raise the pulse detection
threshold. The rolling average baseline calculation is restarted at
the end of the posttrigger samples.

\subsection{\label{sub:Valid-Pulse-Trigger}Valid Pulse Trigger Gate}

Whenever the memory bank on one Struck digitizer fills up, all Struck
modules switch memory banks. This takes up to $290\,\mbox{\ensuremath{\mu}s}$.
This deadtime is not significant if the pulse rate is dominated by
S1 and S2 light from particle interactions in the detector. However,
if one channel is dominated by noise, it can introduce significant
deadtime by forcing the Struck modules to switch memory banks frequently.
This noise may not be coincident across more than one channel. The
Valid Pulse Trigger Gate (VPTG) eliminates this problem. The VPTG
ensures that only those pulses that are coincident in more than one
channel are recorded to memory. It is expected that no S1 or S2 pulses
will occur in only one channel, whereas many types of noise glitches
will.

\begin{figure}
\includegraphics[width=1\columnwidth]{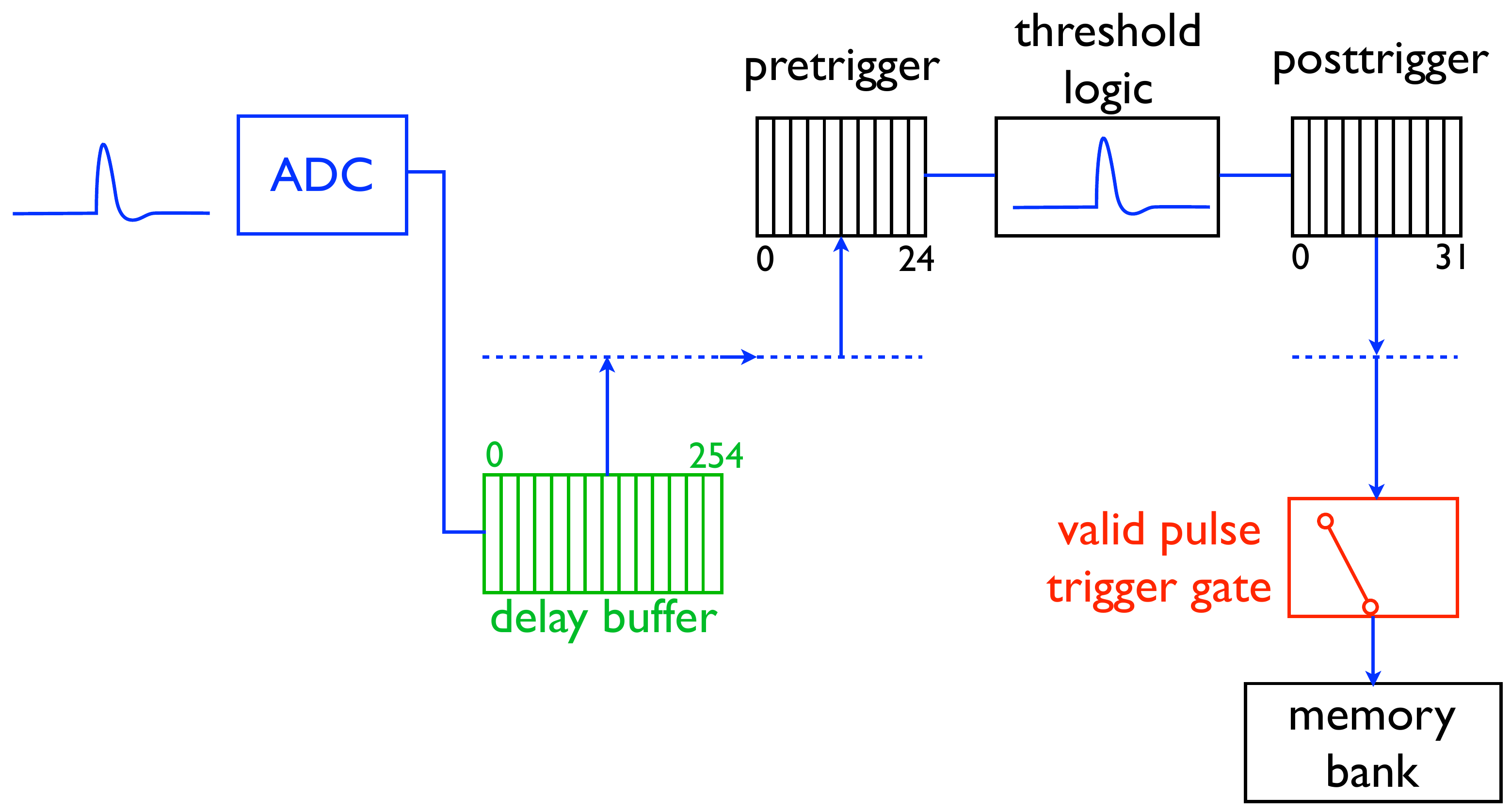}

\caption{\label{fig:VPTG}Diagram of Valid Pulse Trigger Gate (VPTG) and delay
buffer implementation in the Struck ADC.}

\end{figure}

The VPTG is implemented using CAEN V814 Discriminators with majority
coincidence logic. A majority decision is made in about $2\,\mbox{\ensuremath{\mu}s}$,
limited by the width of the S2 pulse. The gate will only turn ON when
more than one channel detects a signal. After the ADC, the signal
is passed through a variable delay buffer between 0 and 254 samples
($0-2.42\,\mbox{\ensuremath{\mu}s}$). This is required because the
Valid Pulse Trigger Gate is delayed by as much as $2\,\mbox{\ensuremath{\mu}s}$
relative to the actual signal. After exiting the delay buffer, the
signal enters the variable length pretrigger region which can be between
0 and 24 samples ($0-240\,\mbox{ns}$). The signal then passes through
the threshold logic, where the beginning and end of the pulse are
determined. Between 0 and 31 posttrigger samples ($0-310\,\mbox{ns}$)
are recorded after the end of the pulse. If the VPTG is ON, then the
data between the start of the pretrigger and the end of the posttrigger
is stored in the memory bank. If the VPTG is OFF, then the pulse is
discarded. This is summarized in Fig. \ref{fig:VPTG}. Using this
delay buffer and the VPTG, it is possible to synchronize the threshold
logic of the Struck with the coincidence decision made by the discriminators. 

\section{DAQ Control}

Each Struck digitizer has four programmable LEMO inputs and four outputs.
These are listed in Table \ref{tab:StruckInOut}.%
\begin{table}
\begin{tabular}{|>{\centering}p{0.02\columnwidth}|>{\centering}m{0.37\columnwidth}|>{\centering}m{0.45\columnwidth}|}
\hline 
 & Inputs & Outputs\tabularnewline
\hline 
\hline 
1 & Clear Timestamp or VPTG & Sample Gate\tabularnewline
\hline 
2 & Stop & Acquisition Stopped or Threshold Crossed\tabularnewline
\hline 
3 & Start & End Address Threshold\tabularnewline
\hline 
4 & Clock In & Clock Out\tabularnewline
\hline 
\end{tabular}

\caption{\label{tab:StruckInOut}Function of Struck SIS3301 control inputs
and outputs.}

\end{table}
The clock and timestamp start time of all Struck boards are synchronized
using the control inputs/outputs. One module serves as the master,
using its own internal clock. Its clock is then provided to all of
the other modules through Clock Out via Philips 740 linear quad fan-out
NIM modules. All other Struck boards are in slave clock mode, and
use the clock provided on their Clock In. This ensures that the clocks
are synchronized across all boards during acquisition. The Clear Timestamp
input is used to synchronize the start of acquisition across all boards.
This signal is generated by the master and provided to the slaves
via LeCroy 128L octal fan-out NIM modules. After acquisition is started,
the function of input 1 is changed via software to the VPTG, which
is discussed in Sect. \ref{sub:Valid-Pulse-Trigger}. Acquisition
on each board can be started and stopped sequentially via software,
or simultaneously via the Start and Stop inputs. The End Address Threshold
output indicates when the board has reached the end of one of its
memory banks and has stopped sampling. Output 2 indicates either that
acquisition has stopped (by software or Stop input) or that the pulse
detection threshold has been crossed and the Struck digitizer is recording
a pulse. The Sample Gate can be used to indicate the livetime of each
board. The gate is ON when the board is sampling and OFF otherwise.
The flexibility of the control inputs/outputs allows for a robust
characterization of the DAQ system.

Data acquisition is initiated by starting all boards via software
and synchronizing their timestamps. When acquisition is stopped on
any one board, indicated by control output 2, all boards are stopped
simultaneously via control input 2. When this happens, the memory
banks are switched in software and the master is started. At this
point the sampling gate (output 1) turns ON and all other boards are
started simultaneously via control input 3. This minimizes the deadtime
of the DAQ system by starting and stopping all boards simultaneously.

\begin{table*}
\begin{centering}
\begin{tabular}{|>{\centering}m{0.85in}|>{\centering}m{0.7in}|>{\centering}m{1.75in}|>{\centering}m{2in}|}
\hline 
Mode & Max. Rate & Description & Benefit\tabularnewline
\hline 
\hline 
Multi-event & $1.5\,\mbox{kHz}$ raw detector event rate & Acquire all pulses, sorting triggered events offline in software. & Debug triggering system; examine detector background offline, irrespective
of trigger.\tabularnewline
\hline 
Single-event & $300\,\mbox{Hz}$ triggered events & Cycle acquisition memory banks after every trigger to download only
pulses in triggered events. & Acquire only triggered events, making full use of digital trigger.\tabularnewline
\hline 
\end{tabular}
\par\end{centering}

\caption{\label{tab:rates}Summary of acquisition modes and their uses.}

\end{table*}

\subsection{Software}

The acquisition software is used to start and stop acquisition, download
and write data to disk, and manage the switching of memory banks.
The acquisition settings are displayed in a graphical user interface
and stored in a MySQL database. The settings are also copied to the
header of each data file as it is written, allowing for easy reference
when loading files. The settings specified include the length of the
acquisition, which channels are enabled, threshold levels, the number
of pretrigger and posttrigger samples, and the number of samples to
average over when calculating the rolling baseline. The software itself
is written in C and runs on a Linux computer with Struck drivers for
communication with the VME crate. The data is written to disk in a
binary format similar to that read out from the Struck memory banks
in order to reduce deadtime due to file writing. The data is further
sorted by the Event Builder, as discussed in Sect. \ref{sec:Data-Formatting-and}.

\subsection{Livetime\label{sub:Livetime}}

When the DAQ software initiates an acquisition, the first memory bank
on each Struck digitizer board is enabled. The initialization timestamp
is recorded for each board whenever a memory bank is enabled. The
acquisition is considered live when all boards are acquiring. When
the acquisition is stopped, the timestamp indicating when the first
board stopped acquiring is recorded. The acquisition is considered
dead when any one board is not acquiring. These values are recorded
in the header of the data file.

\section{\label{sec:Acquisition-Strategies}Acquisition Strategies}

The LUX DAQ operates in either calibration or WIMP search mode. These
modes have very different requirements. Calibration sources are chosen
so that the total event rate in the detector does not exceed $200\,\mbox{Hz}$.
This rate keeps the percentage of events that overlap in time below
10\%, for a maximum drift time of $230\,\mbox{\ensuremath{\mu}s}$.
In WIMP search mode, the expected background rate in the detector
is $1.2\,\mbox{Hz}$ \citep{deViveiros2009thesis}. 

The flexibility of the DAQ design allows for two acquisition strategies
to be used: multi-event mode and single-event mode. They are summarized
in Table \ref{tab:rates} and described in this section.

\subsection{Multi-event Mode}

In multi-event mode, the DAQ records all pulses from the PMTs, regardless
of the trigger condition. Trigger information is recorded as part
of the data stream and used to identify events in software. When the
memory bank of one Struck digitizer becomes full, all of the data
from all Struck digitizers is downloaded to the DAQ computer. The
maximum acquisition rate for multi-event mode is $1.5\,\mbox{kHz}$
without any deadtime due to data transfer from the digitizers. Multi-event
mode also saves pulses in the detector that are not necessarily part
of a triggered event. This allows for the trigger and DAQ systems
to be debugged and for all pulses in the detector to be more closely
examined.

\subsection{Single-event Mode}

In single-event mode, the DAQ alternates between Struck memory banks
for every trigger. Only the pulses that belong to the event that was
flagged by the trigger system are downloaded to the DAQ computer.
This fully utilizes the DDC-8 trigger to pre-select desired events.
This keeps unwanted events from congesting the data stream and analysis.
Because the DAQ needs to cycle Struck memory banks for each event,
the maximum acquisition rate without deadtime is limited to about
$300\,\mbox{Hz}$. Note that this is the rate of events flagged by
the trigger system, not the rate of all particle interactions in the
detector. The latter is limited to $200\,\mbox{Hz}$ by event overlap
constraints. The ability to pre-select events using the DDC-8 makes
this mode ideal for acquiring calibration data because it is desirable
to record only events with certain energies or at certain locations
within the active region.

\section{\label{sec:Data-Formatting-and}Data Formatting and Reduction}

When the raw data from the Struck digitizers is written to file, it
is sorted by channel and then by pulse. The sorting is chronological,
but shows no relation to the actual event triggers. The Event Builder
uses the trigger information to sort the data by event and then by
channel, and finally by pulse. An index file is also recorded to easily
relate the location of each event in the event files. The Event Builder
serves three purposes: (1) it sorts the data so that it is easily
parsed for event traces; (2) it incorporates traces from the water
shield PMTs as well as information from the trigger system into the
data stream; and, (3) it discards any pulses that were recorded in
the file that do not belong to an event (lying in the event window
determined by the trigger mode). Item (3) is particularly useful in
multi-event mode where all pulses recorded by the Struck digitizers
are written to file, independent of the event triggers. The Event
Builder is written in C and runs in parallel to the acquisition on
the DAQ computer. The Event Builder writes one output file per raw
data file, and one index file per acquisition. The number of events
in a raw file can be specified at the start of the acquisition to
keep size and number of files from becoming cumbersome. Acquisition
settings, Event Builder settings, trigger settings, and water tank
readout settings are all recorded in the header of each file.

\section{Conclusion}

The LUX data acquisition system is finely tuned and optimized for
the LUX dark matter experiment. This is achieved mainly by using custom-built
analog electronics and custom digitization firmware. The dynamic range
is optimized for performance at low energies for dark matter searches,
as well as for high-energy calibrations. The noise level is limited
such that 95\% of single photoelectrons are clearly visible above
a $5\sigma$ fluctuation in the baseline noise. A novel real-time
baseline suppression technique has been employed in the digitizers
to provide a factor of $\times1/50$ reduction in storage size. The
maximum event rate achieved by the DAQ, without incurring significant
deadtime, is $1.5\,\mbox{kHz}$. This is well above what is needed
for calibrations and WIMP searches. While being finely tuned, the
LUX DAQ maintains flexibility in acquisition techniques and strategies.

\section{Acknowledgments}

The authors would like to recognize the assistance of Matthias Kirsch and Tino Haeupke at Struck Innovative Systeme for their work on the ADC firmware. This work was partially supported by the U.S. Department of Energy (DOE) grants DE-FG02-08ER41549, DE-FG02-91ER40688, DE-FG02-95ER40917, DE-FG02-91ER40674, DE-FG02-\newline11ER41738, DE-FG02-11ER41751, DE-AC52-07NA27344, the U.S. National Science Foundation grants PHYS-0750671, PHY-0801536, PHY-1004661, PHY-1102470, PHY-1003660, the Research Corporation grant RA0350, the Center for Ultra-low Background Experiments at DUSEL (CUBED), and the South Dakota School of Mines and Technology (SDSMT). We gratefully acknowledge the logistical and technical support and the access to laboratory infrastructure provided to us by the Sanford Underground Research Facility (SURF) and its personnel at Lead, South Dakota.

\bibliographystyle{utcaps2.3_LdV}
\bibliography{LUX_BibTeX_v1}

\end{document}